\documentclass[prb,twocolumn,floatfix]{revtex4}
\usepackage{graphicx}
\usepackage[utf8]{inputenc}
\usepackage{mathtools}
\usepackage{amsmath}
\usepackage{comment}
\usepackage{hyperref}
\hypersetup{
  colorlinks,
  citecolor=magenta,
  linkcolor=blue,
  urlcolor=blue}
\usepackage{bbm}
\usepackage{subfigure}
\usepackage{makecell}
\begin{document}

\title{
First order phase transitions in the square lattice ``easy-plane'' J-Q model}

\author{Nisheeta Desai}
\affiliation{Department of Physics \& Astronomy, 
University of Kentucky, 
Lexington, KY 40506.}
\author{Ribhu K. Kaul}
\affiliation{Department of Physics \& Astronomy, 
University of Kentucky, 
Lexington, KY 40506.}

\begin{abstract}
We study the quantum phase transition between the superfluid and valence bond solid in ``easy-plane" J-Q models on the square lattice. The Hamiltonian we study is a linear combination of two model Hamiltonians: (1) an SU(2) symmetric model, which is the well known J-Q model that does not show any direct signs of a discontinuous transition on the largest lattices and is presumed continuous, and (2) an easy plane version of the J-Q model, which shows clear evidence for a first order transition even on rather small lattices of size $L\approx16$. A parameter $0\leq\lambda\leq 1$ ($\lambda=0$ being the easy-plane model and $\lambda=1$ being the SU(2) symmetric J-Q model) allows us to smoothly interpolate between these two limiting models.  We use stochastic series expansion (SSE) quantum Monte Carlo (QMC) to investigate the nature of this transition as $\lambda$ is varied - here we present studies for $\lambda=0,0.5,0.75,0.85,0.95$ and $1$. While we find that the first order transition weakens as $\lambda$ is increased from 0 to 1, we find no evidence that the transition becomes continuous until the SU(2) symmetric point, $\lambda=1$. We thus conclude that the square lattice superfluid-VBS transition in the two-component easy-plane model is generically first order. 
\end{abstract}

\maketitle

\section{Introduction}
\label{sec:intro}

The quantum transition from N\'eel or superfluid to a valence bond solid (VBS) has been proposed to be described by the deconfined criticality scenario.\cite{senthil2004:science,senthil2004:prb} In this scenario it is generically possible to have a direct continuous N\'eel-VBS transition. A number of field theoretic formulations that describe this putative critical point at long distances have been put forward and interesting connections between different representations have been conjectured via duality arguments.~\cite{senthil2006:topnlsm, tanaka2005:nvbs,senthil2017:dualities} Establishing these fascinating connections non-perturbatively by lattice simulations is an exciting field of current research. In the original study two kinds of symmetries were highlighted for their possibility as platforms for deconfined criticality, an SU(2) symmetric system and a U(1)$\times$Z$_2$ symmetric system. Physically, the SU(2) field theory could be a description for a rotationally symmetric $S=1/2$ anti-ferromagnet and its transition to a valence bond solid. The U(1)$\times$Z$_2$  system can be thought of as a model for the same N\'eel-VBS transition in  magnet with easy-plane anisotropy or alternatively as a model for a superfluid to Mott transition. \footnote{We will use the terms magnetic and superfluid interchangeably throughout this manuscript} 

In the years since the original proposal, it has been demonstrated that the N\'eel-VBS transition and many of its variants can be studied in sign problem free quantum spin Hamiltonian models on large lattices.\cite{kaul2013:qmc} Through extensive numerical simulations in the SU(2) symmetric models many aspects of the proposal have been borne out and no direct evidence for a first order transition has been observed. \cite{sandvik2007:deconf,melko2008:fan,lou2009:q3model,sandvik2010:deconf,banerjee2010:prb,pujari2013:prl,harada2013:suNHeis,pujari2015:prb,Shao2016:science,shao2017:nearlydeconf,yao2019:arxiv} Numerical studies of classical statistical mechanics models of tightly packed loops and dimer models in three dimensions that have been argued to realize the same universal physics as the SU(2) N\'eel-VBS transition are also consistent with the deconfined criticality scenario. \cite{nahum2015:deconfloop,powell2008:prl,powell2009:prb,charrier2008:prl,chen2009:prb,sreejith2014:prb,alet2006:prl} Despite this large body of evidence for the deconfined criticality scenario, numerical studies have observed scaling violations whose origin is currently unclear.\cite{kaul2011:su34,nahum2015:deconfloop,jiang2008,kuklov2006:sciencedirect,Shao2016:science}

\begin{figure}[t]
 \includegraphics[width=0.5\textwidth]{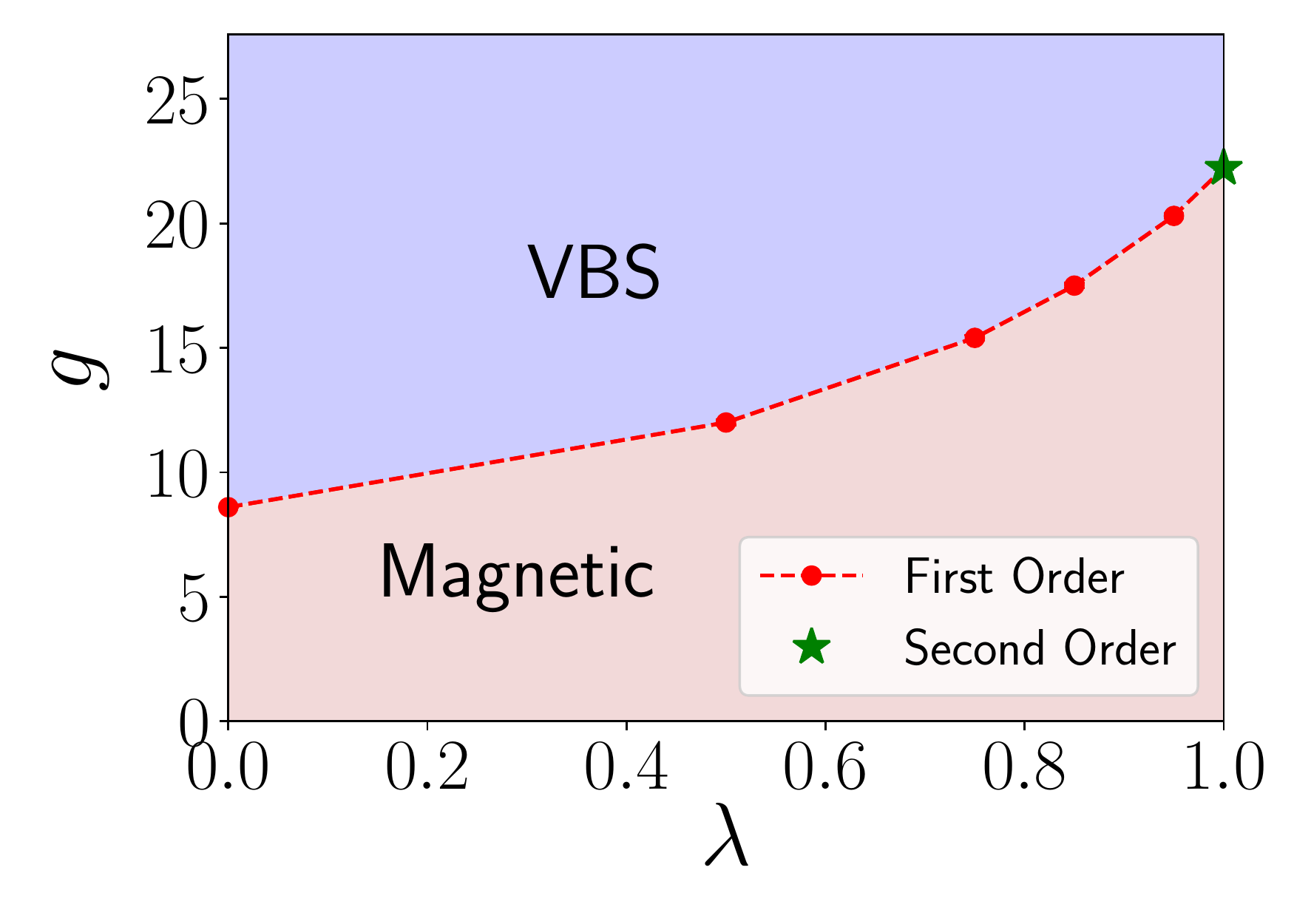}
 \caption{\label{fig:phasediag} Phase Diagram of $H^{JQ}_{\lambda}$ described by Eq. \ref{eq:ham} as a function of $\lambda$ and $g\equiv Q/J$. 
 Using the model $H^{JQ}_{\lambda}$ we can access the phase boundary between the N\'eel and VBS phases. The transition at $\lambda=0$ was demonstrated to be first order previously.~\cite{jon2016:easyplane} We find that this transition is first order for all values of $\lambda<1$. The signals of first order behavior that we detect in our QMC simulations vanish at the symmetric point $\lambda=1$ even on the largest lattices simulated here. We note that when $\lambda=1$ the model has a higher SU(2) symmetry, everywhere else in this phase diagram it only has the generic U(1)$\times Z_2$ ``easy-plane" symmetry.}
\end{figure}

In  contrast in the easy-plane case where SU(2) is broken to U(1)$\times Z_2$ a number of numerical studies have concluded that the transition is first order.\cite{sandvik2002:prl,kragset2006:prl,kuklov2006:sciencedirect,sen2007:prb,jon2016:easyplane,jon2017:prl} 
Recently however it has been claimed that a continuous transition has been found in a square lattice model with somewhat weaker easy-plane anisotropy,\cite{meng2017:dualities,ma2018:prb} suggesting that perhaps a large easy-plane anisotropy could result in a first order transition, and the first order and second order regime are separated by a multicritical point.\footnote{We note for completeness that a model of hard core bosons at 1/3 filling on the Kagome lattice has been argued to be described by a similar field theory and host a putative easy-plane deconfined critical point.\cite{zhang2018:kagome} Although it was originally determined to have a first order transition,\cite{isakov2006:prl} it has been claimed in recent work to host a continuous transition.\cite{zhang2018:kagome}}
Motivated by this study, we address the issue of how the easy-plane transition is connected to the symmetric one,  by studying a model that interpolates between these two limiting cases on the square lattice. For the symmetric model we use the popular J-Q model which shows no direct evidence for first order behavior even on the largest studied lattice sizes. For the easy plane case we use an easy-plane J-Q that was shown to have a first order transition already visible on $L\approx 16$. The interpolating model introduced in detail below is slightly different from the one studied in Ref.~\onlinecite{meng2017:dualities,ma2018:prb} where the easy-plane anisotropy was introduced only in the J-term; both models are believed to have the same universal features however. In this work we present studies on larger lattices and a more thorough analysis. Contrary to the previous study, we find no evidence for new continuous easy-plane criticality. Instead we find a first order transition for $0\leq \lambda <1$ that weakens as $\lambda$ is increased and we approach the symmetric point ($\lambda=1$) at which all our direct signals of a first order transition vanish and the transition is presumed continuous. This is the primary result of our paper and is summarized in Fig.~\ref{fig:phasediag}. Although no numerical study can rule out that the transition becomes continuous for a very small but finite window close to $\lambda=1$ (with finite easy-plane anisotropy), we find this rather unlikely given our results below. We thus conclude that the easy-plane N\'eel-VBS transition is generically first order on the square lattice.

\section{The Model}
\label{sec:model}
\begin{figure}[!t]
 \includegraphics[width=0.5\textwidth]{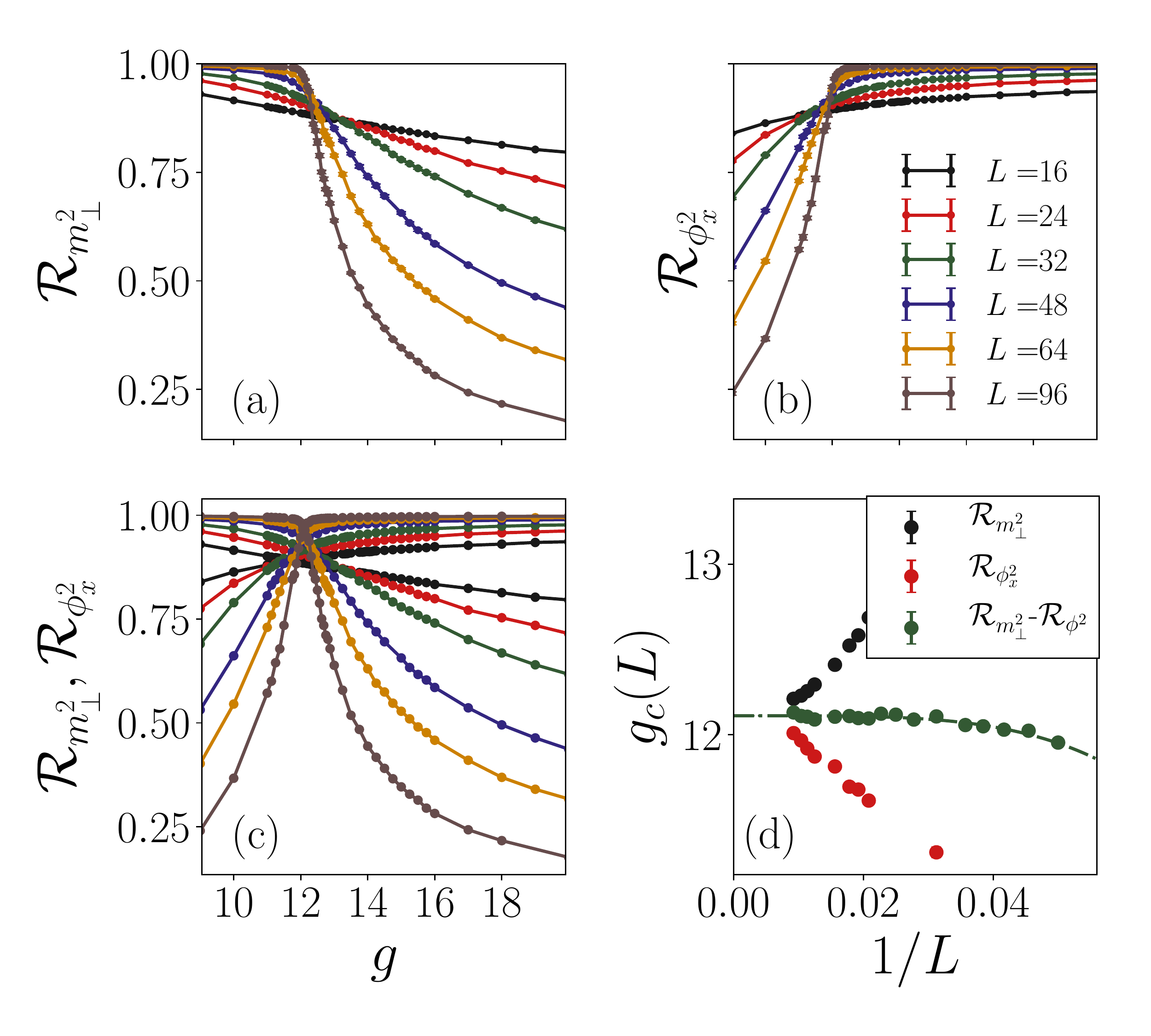}
  \caption{Quantum Monte Carlo results for the transition from superfluid to the VBS phase for $\lambda=0.5$: (a)-(b) show the quantities $\mathcal{R}_{m_{\perp}^2}$ and $\mathcal{R}_{\phi_x^2}$ respectively, defined by Eq.~(\ref{eq:ratios}) cross for different $L$ at the transition point. The $\hat x$-axis on these graphs is identical to the one show in (c). (c) The same data as shown in (a-b) but here together, suggesting an accurate estimate for the critical coupling can be obtained from the crossing of $\mathcal{R}_{m_{\perp}^2}$ and $\mathcal{R}_{\phi_x^2}$ for a given value of $L$. (d) Values of coupling at the crossings of $L$ and $L/2$, for $\mathcal{R}_{m_{\perp}^2}$ and $\mathcal{R}_{\phi_x^2}$ are plotted vs $1/L$. $g_c(L)$ from crossing analysis of $\mathcal{R}_{m_{\perp}^2}$-$\mathcal{R}_{\phi_x^2}$ as suggested in (c) is shown to fit to a form $g_c(L)=g^*_c+\frac{C}{L^{e}}$ where $g^*_c=12.111(3)$. This fitting has been done for $L\leq 64$ since larger sizes deviate from this fitting form. We demonstrate in Fig.~\ref{fig:histlam50} that this deviation arises due to the formation of double peaks in the histograms for the Monte Carlo estimators, a classic sign of a first order transition.}
\label{fig:ratioslam50}
\end{figure}

The Hamiltonian studied here is an $S=\frac{1}{2}$ system on an $L \times L$ square lattice,
\begin{equation}
 H^{JQ}_{\lambda}=\lambda H^{JQ}_s + (1-\lambda)H^{JQ}_{ep},
 \label{eq:ham}
\end{equation}
 and is a linear combination of two parts, $H^{JQ}_s$ is the SU(2) symmetric part and $H^{JQ}_{ep}$ is the \textit{easy plane} part that explicitly breaks the SU(2) symmetry. $\lambda$ is an anisotropy 
parameter that allows us to smoothly interpolate between the easy plane limit ($\lambda=0$) and the SU(2) symmetric limit ($\lambda=1$). We define the 
singlet projection operator on a bond between two sites $i$ and $j$ as,
\begin{equation}
 P_{ij}= \frac{1}{4}- \vec{S_i}.\vec{S_j}.
 \label{eq:singproj}
\end{equation}
$S^{\mu}_i=\frac{1}{2} \sigma^{\mu}_i$, are standard spin-$\frac{1}{2}$ operators where $\sigma^{\mu}_i$ are Pauli matrices.

Then $H^{JQ}_s$, which is the well known J-Q model, \cite{sandvik2007:deconf} can be written as, 
\begin{equation}
  H^{JQ}_s = -J\sum_{\langle ij \rangle} P_{ij} - Q \sum_{\langle ijkl \rangle} P_{ij} P_{kl}
\label{eq:JQ}
\end{equation}
The second term in the above equation is a sum over all elementary plaquettes $ijkl$. $H^{JQ}_s$ has full SU(2) symmetry inherited from $P_{ij}$. Similarly if we define 
\begin{equation}
 \tilde{P}_{ij}= S^x_iS^x_j+S^y_iS^y_j
 \label{pijtilde}
\end{equation}
the easy plane Hamiltonian, $H^{JQ}_{ep}$ can be written as,\cite{jon2016:easyplane}
\begin{equation}
 H^{JQ}_{ep} = J\sum_{\langle ij \rangle} \tilde{P}_{ij}  - Q \sum_{\langle ijkl \rangle} \tilde{P}_{ij}\tilde{P}_{kl} 
\label{eq:epJQ}
\end{equation}
$\tilde{P}_{ij}$ has a symmetry of U(1) $\times$ Z$_2$, which corresponds to U(1) rotations about the $\hat z$-axis and the Z$_2$ operation of a $\pi$ rotation about the $\hat x$-axis.

We study the quantum phase transition from the magnetic phase to the valence-bond solid (VBS) phase as $g \equiv Q/J$ is varied for a fixed $\lambda$. While in the easy plane limit, i.e. $\lambda=0$, this transition has been found to be first order, \cite{jon2016:easyplane} it has been argued to be to continuous in the SU(2) symmetric limit, $\lambda=1$. \cite{sandvik2007:deconf,melko2008:fan,sandvik2010:deconf,harada2013:suNHeis} In this work we interpolate between the two limiting models with the aim of elucidating the evolution of the nature of the quantum transition and in particular to investigate whether the transition is continuous for any $\lambda<1$.

\section{Numerical Simulations}
\label{sec:results}

The numerical results presented below have been obtained using the stochastic series expansion (SSE) quantum Monte Carlo method. \cite{sandvik2010:vietri} We use the directed loop algorithm \cite{directed-loops} to carry out global loop updates on our Monte Carlo configurations (see Appendix \ref{subsec:ham}).

\begin{figure}[t]
  \begin{minipage}[b]{0.9\linewidth}
    \centering
    \includegraphics[width=0.9\linewidth]{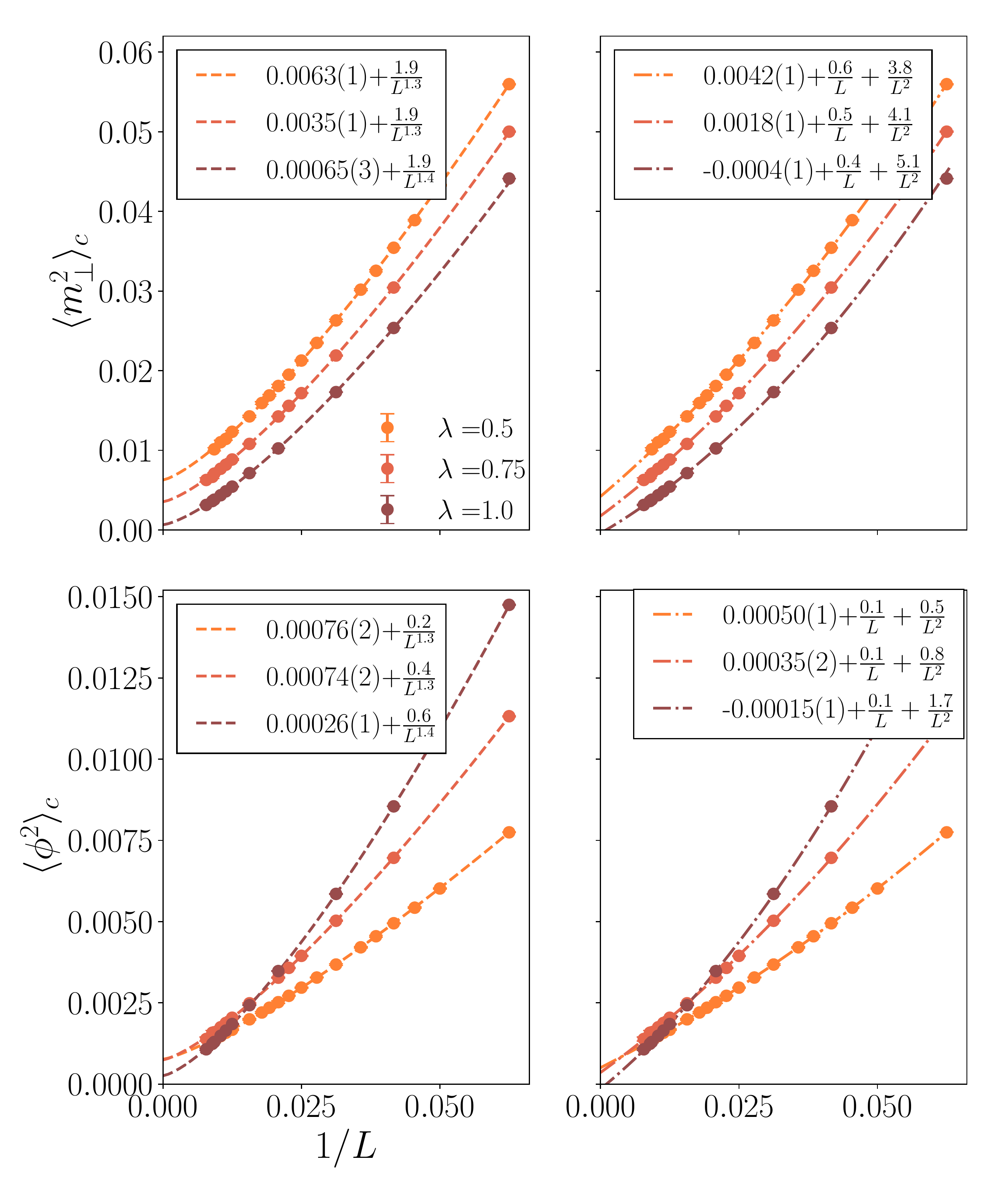}
  \end{minipage}
  \vspace*{1cm}
  \begin{minipage}[b]{0.75\linewidth}
    \centering
    \begin{tabular}{|c|c|c|c|c|}
        \hline
        \multicolumn{5}{|c|}{$\chi^2$ per degree of freedom} \\
        \hline \hline
        $\lambda$ & \thead{$m^2$ \\ Power law} & \thead{$m^2$ \\ Polynomial} & \thead{$\phi^2$ \\ Power law} & \thead{$\phi^2$ \\ Polynomial} \\ \hline
         0.5 & 0.43 & 0.4 & 0.17 & 0.23 \\ \hline
         0.75 & 1.3 & 1.0 & 0.42 & 0.89 \\ \hline
         1.0 & 0.53 & 1.8 & 1.25 & 0.69 \\ 
        \hline
    \end{tabular}
\end{minipage}
\caption{Finite-size scaling of the superfluid and VBS order parameters at the critical point with extrapolations to the thermodynamic limit for $\lambda=0.5,0.75$ and 1. Shown are  $\langle m_{\perp}^2 \rangle$ and $\langle \phi^2_x \rangle$ as a function of $L$ evaluated at the finite size pseudo-critical coupling $g_c(L)$. These couplings $g_c(L)$ are determined by estimating where $\mathcal{R}_{m^2}$ and $\mathcal{R}_{\phi^2}$ for the same value of $L$ cross each other as described in Fig.~\ref{fig:ratioslam50}. Dashed lines show extrapolations of the finite size data. Extrapolations have been carried out for two different fit forms, (a) Power law: $C_0+\frac{C_1}{L^{e_1}}$ (left) (b) Polynomial: $C_0+\frac{C_1}{L}+\frac{C_2}{L^2}$ (right). The value of $\chi^2$ per degree of freedom for these fits is shown in the table below the figure, indicating the reliability of the fits. The biggest system size used for the fits is $L=128$. We find that the numerical values to which $\langle m^2_{\perp} \rangle_c$ and $\langle \phi^2_{x} \rangle_c$ extrapolate depend on the fit form itself and are inconsistent with the stochastic errors (shown in the legend). In both fit forms the extrapolated order parameters go unambiguously to finite values for $\lambda=0.5$ and $\lambda=0.75$. For $\lambda=1$ on the other hand they are consistent with a zero extrapolated value. The $\langle m_{\perp}^2 \rangle$ data shows this effect much more clearly than in the $\langle \phi^2_x \rangle$, where it is nonetheless also evident.}
\label{fig:ordrprmgcextrpln}
\end{figure}

\begin{figure}[t]
  \begin{minipage}[b]{0.9\linewidth}
    \centering
    \includegraphics[width=0.9\linewidth]{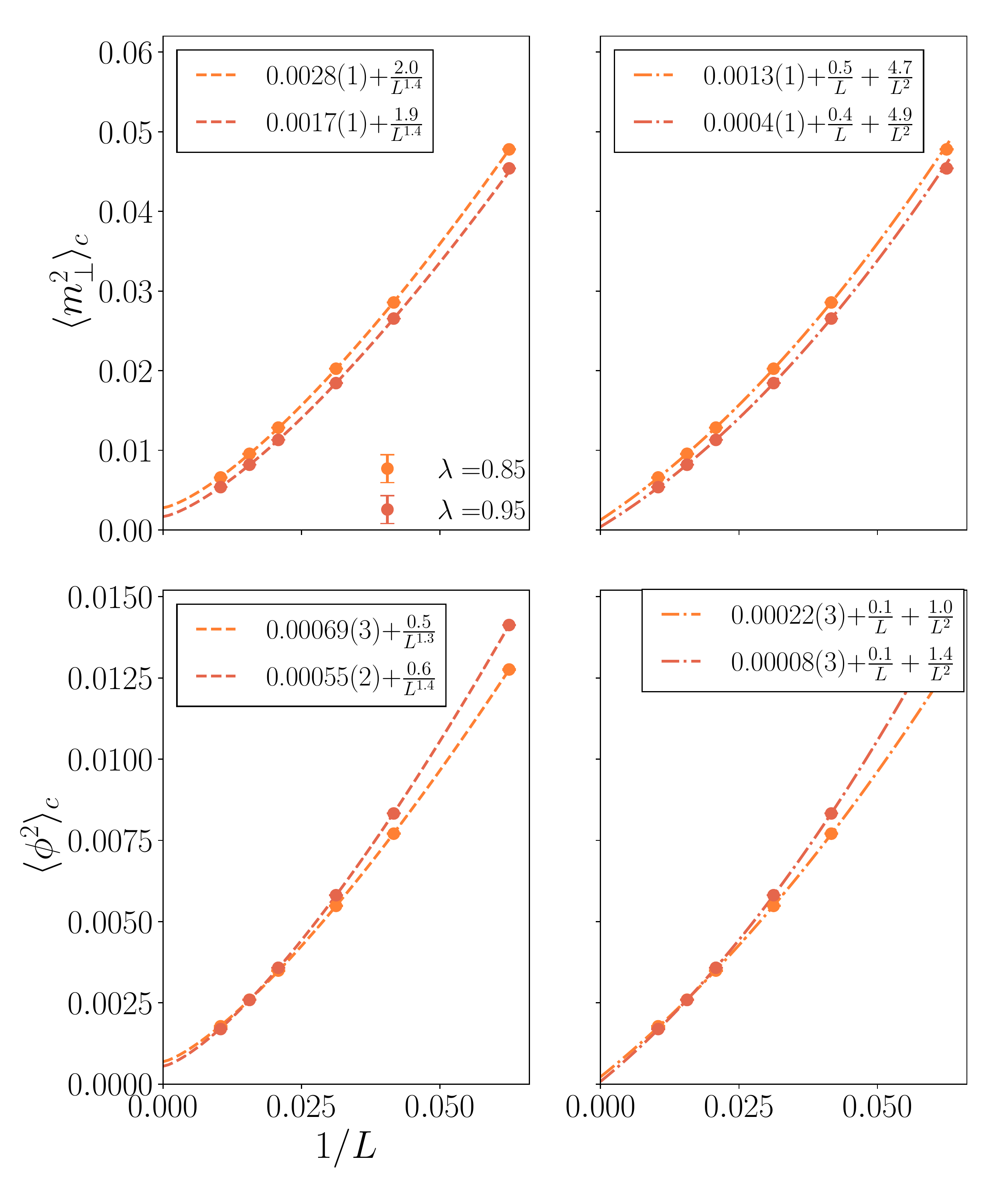}
  \end{minipage}
  \vspace*{1cm}
  \begin{minipage}[b]{0.9\linewidth}
    \centering
    \begin{tabular}{|c|c|c|c|c|}
        \hline
        \multicolumn{5}{|c|}{$\chi^2$ per degree of freedom} \\
        \hline \hline
        $\lambda$ & \thead{$m^2$ \\ Power law} & \thead{$m^2$ \\ Polynomial} & \thead{$\phi^2$ \\ Power law} & \thead{$\phi^2$ \\ Polynomial} \\ \hline
         0.85 & 1.96 & 0.79 & 0.71 & 1.23 \\ \hline
         0.95 & 0.62 & 2.87 & 2.4 & 1.35 \\
        \hline
    \end{tabular}
\end{minipage}
\caption{\label{fig:ordrprmgcextrpln2}This is the same kind of analysis as fig.~\ref{fig:ordrprmgcextrpln} but now for $\lambda=0.85,0.95$. We have shown these two values of $\lambda$  separately to avoid overcrowding. The largest value of $L$ used for this analysis is $L=96$. It can be inferred that both the order parameters extrapolate to a finite value for $\lambda=0.85$ in the thermodynamic limit using both the fitting forms. For $\lambda=0.95$ they clearly extrapolate to a finite value in the power law fit. However, they extrapolate to a very small positive value in the polynomial fit which is within four times the error bar, therefore we are unable to draw a very reliable conclusion here. Fig~\ref{fig:ordrprmcvslam} shows the evolution of the extrapolated quantities from Figs.~\ref{fig:ordrprmgcextrpln} and~\ref{fig:ordrprmgcextrpln2}  as a function of $\lambda$ for $\lambda=0.5,0.75,0.85,0.95,1.0$.}
\end{figure}

Fig. \ref{fig:phasediag} shows a phase diagram obtained from numerical simulations as a function of the coupling $g=Q/J$ and the anisotropy parameter $\lambda$. For a given $\lambda$, on increasing $g$ we find a quantum phase transition from the magnetic to the VBS phase. Here, we work in units where $\sqrt{J^2+Q^2}=1$ and at an inverse temperature $\beta=L$ for an $L\times L$ lattice. All data presented has been tested to be in the $T=0$ limit as demonstrated in Appendix \ref{subsec:obsvsbeta}. \\

\subsection{Measurements}
\label{subsec:measurements}

When $\lambda < 1$, the presence of a small amount of anisotropy makes the spins preferentially align in the $XY$ plane. Therefore as we vary $g$ in our simulations, we look for a phase transition between the $XY$ order (superfluid) and VBS phases. We define the following quantities to detect magnetic order,

\begin{equation}
    S_{m_{\perp}^2}(\vec{k}) = \sum_{\vec{r}}  e^{i \vec{k}.\vec{r}} \langle S^x_{\vec{0}}S^x_{\vec{r}}+S^y_{\vec{0}}S^y_{\vec{r}}\rangle
    \label{eq:m2perpcorr}
\end{equation}

\begin{equation}
    S_{m_{\parallel}^2}(\vec{k}) = \sum_{\vec{r}}  e^{i \vec{k}.\vec{r}} \langle S^z_{\vec{0}}S^z_{\vec{r}}\rangle
    \label{eq:m2parcorr}
\end{equation}

The square of the superfluid order parameter is $\langle m_{\perp}^2\rangle =S_{m_{\perp}^2}(\pi,\pi)$. The quantity $\langle S^x_{\vec{0}}S^x_{\vec{r}}+S^y_{\vec{0}}S^y_{\vec{r}}\rangle$ is measured during the loop update by keeping track of the distance between the head and the tail of the loop when they are at the same time slice. \cite{dorneich2001:pre} We also define a N\'eel order parameter square as $\langle m_{\parallel}^2 \rangle=S_{m_{\parallel}^2}(\pi,\pi)$. When $\lambda=1.0$, $S^x$, $S^y$ and $S^z$ are equivalent and therefore $\langle m^2_{\perp} \rangle$ and $\langle m^2_{\parallel} \rangle$ are equal upto a normalization. 

VBS order is one where each spin forms a singlet with its neighbour and the pattern of singlets forms columnar order. We construct the VBS order parameter square from the following quantity,
\begin{equation}
    S_{\phi^2_x}(\vec{k}) = \sum_{\vec{r}} e^{i \vec{k}.\vec{r}} \langle \mathcal{Q}_x(\vec{0})\mathcal{Q}_x(\vec{r}) \rangle.
    \label{eq:plaqcorr}
\end{equation}
Here $\mathcal{Q}_x(\vec{r})$ is a plaquette operator which equals the sum of all the operators in the Hamiltonian acting on the plaquette at $\vec{r}$ as described in Appendix \ref{sec:numsim}. \\

The spin stiffness $\rho_s$ is defined as,
\begin{equation}
  \rho_s=\frac{\partial^2 E(\phi)}{\partial \phi^2}\bigg|_{\phi=0} \\
  \label{stiff}
 \end{equation}
Here E($\phi$) is the energy of the system with a twist of $\phi$ in the boundary condition in either the $x$ or the $y$ direction. In the QMC, this quantity is related to the winding number of loops in the direction that the twist has been added,~\cite{sandvik1997:prb}
\begin{equation}
  \rho_s= \frac{\langle W^2 \rangle}{\beta}\\
  \label{winding}
 \end{equation}
 where $\beta$ is the inverse temperature. $\rho_s$ goes to a finite value in the magnetically ordered phase but goes to 0 otherwise. The quantity $L\rho_s$ is expected to show a crossing for different values of $L$ at the coupling at which magnetic order is destroyed.
 
In order to detect the ordered phase we make use of ratios defined as, 
\begin{equation}
 \mathcal{R}_{o_p}=1-\frac{|S_{o_p}(k'_{o})|}{|S_{o_p}(k_o)|}.
 \label{eq:ratios}
\end{equation} 

Here ${o_p}=m_{\perp}^2,m_{\parallel}^2,\phi^2$; $k_o$ and $k'_o$ are the ordering momentum and momentum closest to the ordering momentum respectively. In the ordered phase $\mathcal{R}_{o_p}$ goes to 1 and in the disordered phase it goes to 0 on increasing system size, therefore they are expected to cross for different system sizes at the critical point. 

\subsection{Numerical Results}
\label{subsec:results}
\begin{figure}[t]
    \centering
    \includegraphics[width=0.5\textwidth]{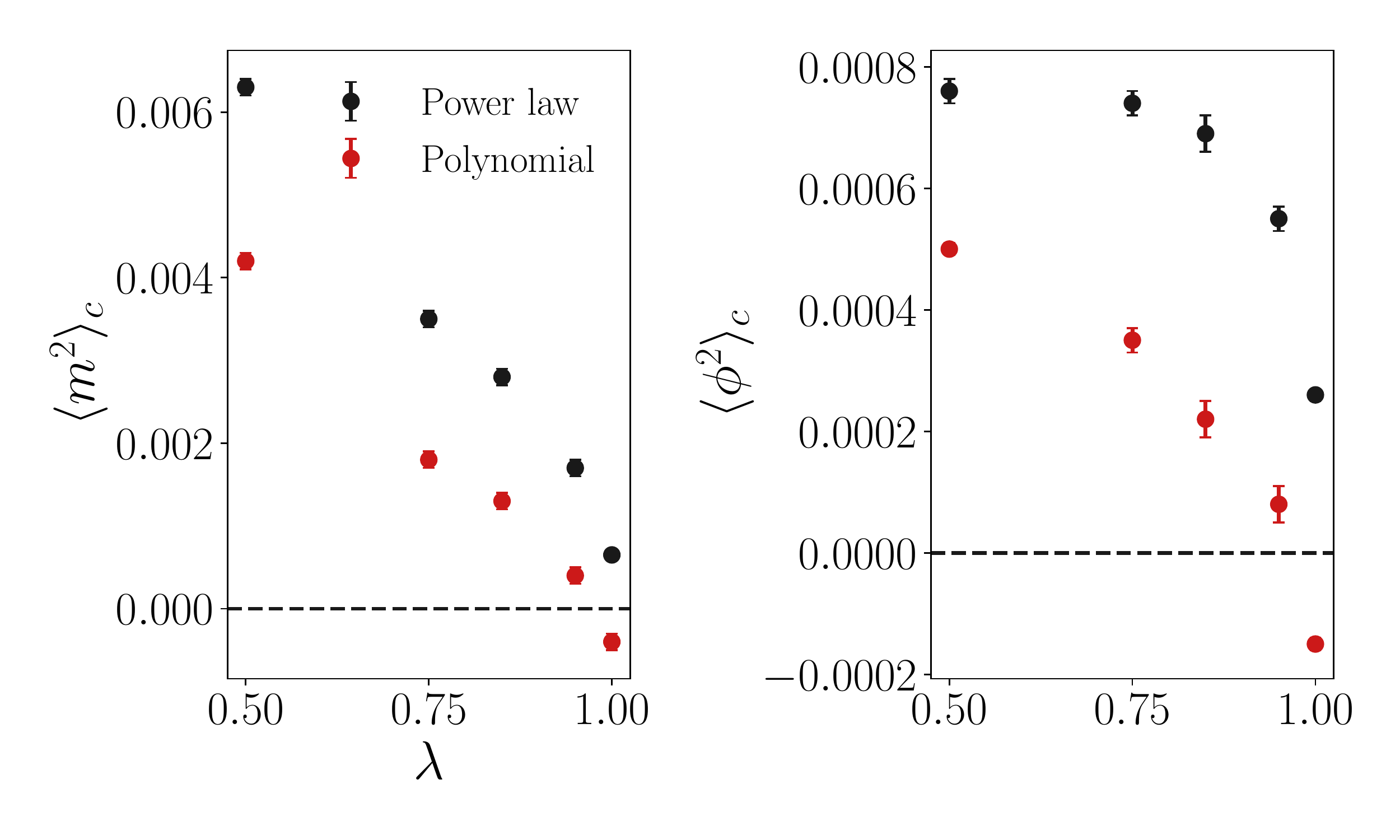}
    \caption{$\langle m^2 \rangle_c$ and $\langle \phi^2 \rangle_c$ as found from extrapolations in  figs~\ref{fig:ordrprmgcextrpln} and~\ref{fig:ordrprmgcextrpln2} using the power law and polynomial fitting forms  plotted as a function of $\lambda$. The error bars shown are stochastic errors, there are in addition systematic errors associated with the extrapolating function used. To estimate the systematic error we note that power law form overestimates and the polynomial underestimates the extrapolated order parameters, they thus provide a window for the order parameters in the thermodynamic limit.  For all $\lambda < 1$ both N\'eel and VBS order parameters extracted from both fit forms are positive indicating that both order parameters are finite at the transition: the transition is hence of first order.  The extrapolated order parameters can be seen to approach zero as $\lambda \rightarrow 1$ (this is smoother for $\langle m^2 \rangle_c$ given the larger values compared to $\langle \phi^2 \rangle_c$). This indicates that the first order transition continuously evolves to a second order transition as $\lambda$ approaches 1. For $\lambda=1$ we find that polynomial extrapolation gives a negative value and the power law gives a positive value consistent with the most extensive studies that find a continuous transition with SU(2) symmetry.~\cite{Shao2016:science}
    \label{fig:ordrprmcvslam}}
\end{figure}{}

\begin{figure}[t]
  \centering
  \includegraphics[width=0.5\textwidth]{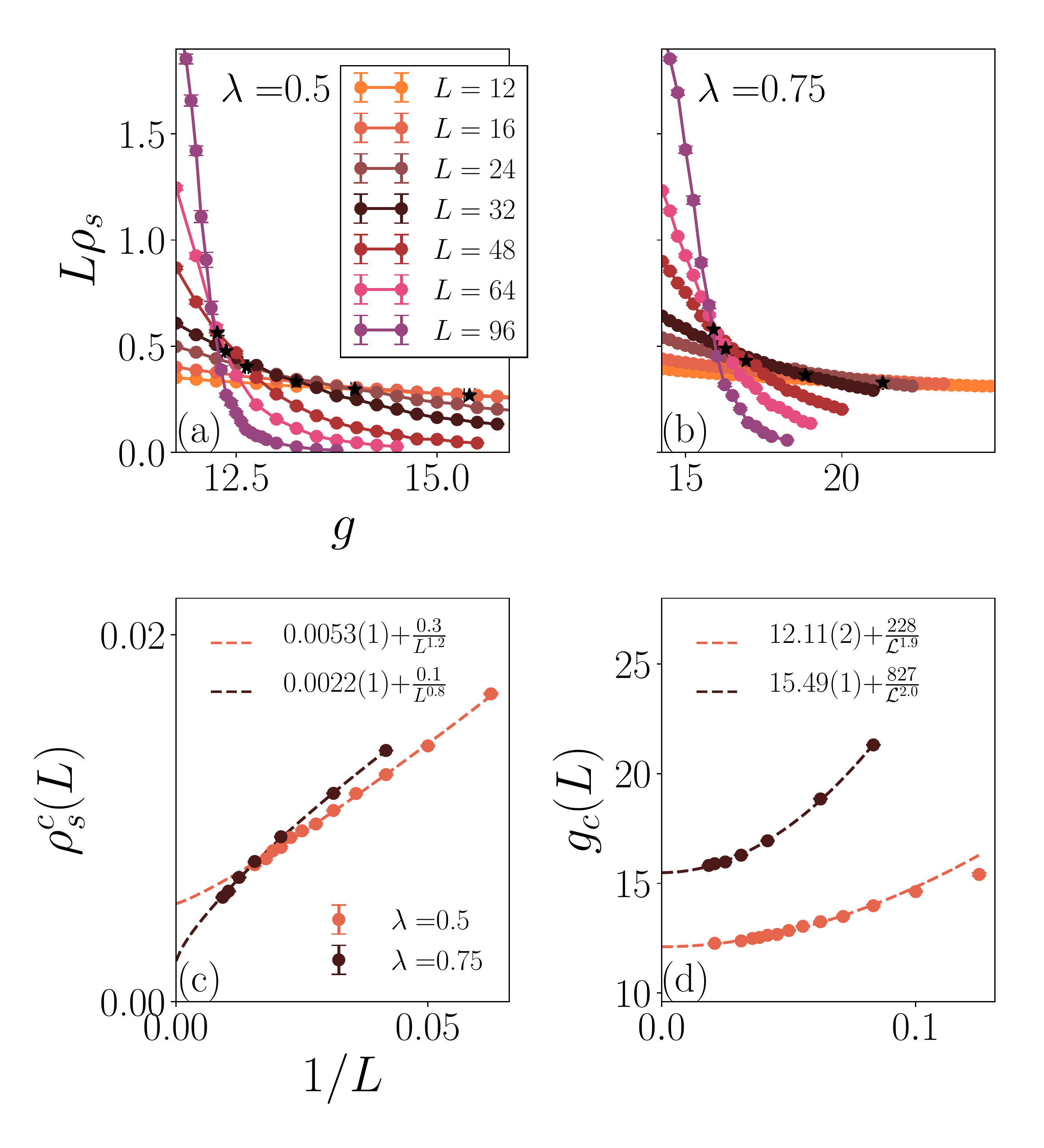}
  \caption{(a)-(b) Crossings of $L \rho_s$ for $\lambda=0.5$ and $\lambda=0.75$ indicating a transition from a magnetic to non-magnetic phase. The black stars denote points where the curves of $L$ and $L/2$ cross. (c) $\rho_s$ extracted at these crossing points is fit to a power law and is shown to extrapolate to a finite value in the thermodynamic limit for both $\lambda=0.5$ and $\lambda=0.75$ (the same analysis for $\lambda=0.85,0.95$ has been shown in fig.~\ref{fig:rhocfit2}). (d) The value of the coupling $g$ at these crossing points, $g_c(L)$, is shown to extrapolate to $g^*_c=12.11(2)$ and $g^*_c=15.49(1)$}
  \label{fig:rhocfit}
 \end{figure}
 
In this work, we focus on four values of the anisotropy parameter, $\lambda=0.5$, $\lambda=0.75$, $\lambda=0.85$ and $\lambda=0.95$. We have included a comparison with the symmetric case $\lambda=1$ when appropriate.

\subsubsection{Crossing Analysis}
\label{subsubsec:crossings}
Fig. \ref{fig:ratioslam50}~(a) and (b) show ratios $\mathcal{R}_{m_{\perp}^2}$ and $\mathcal{R}_{\phi_x^2}$  (defined above) crossing for different $L$ for $\lambda=0.5$. This indicates a transition from the magnetic to VBS phase. Fig. \ref{fig:ratioslam50}(c) shows crossing of these ratios for the same $L$.  As shown in \ref{fig:ratioslam50}(d), the crossing analysis from \ref{fig:ratioslam50}(c) yields the transition point to be at $g^*_c=12.111(3)$, which is close to the value at which the couplings at the crossing points, $g_c(L)$, converge. This extrapolation has been done only using small system sizes, $L\leq 64$. We notice that smaller system sizes can be seen to smoothly converge to $g^*_c\approx12.1$, bigger system sizes start deviating from this trend. This is because of the double peaked structure that starts to develop in the order parameter estimators, making it difficult to reliably extrapolate $g_c(L)$ using bigger lattices. Fig. \ref{fig:rhocfit} shows crossings of the quantity  $L\rho_s$ for both $\lambda=0.5$ and $\lambda=0.75$, which also indicates transition out of the magnetic phase.

\begin{figure}[t]
  \centering     
 \includegraphics[width=0.5\textwidth]{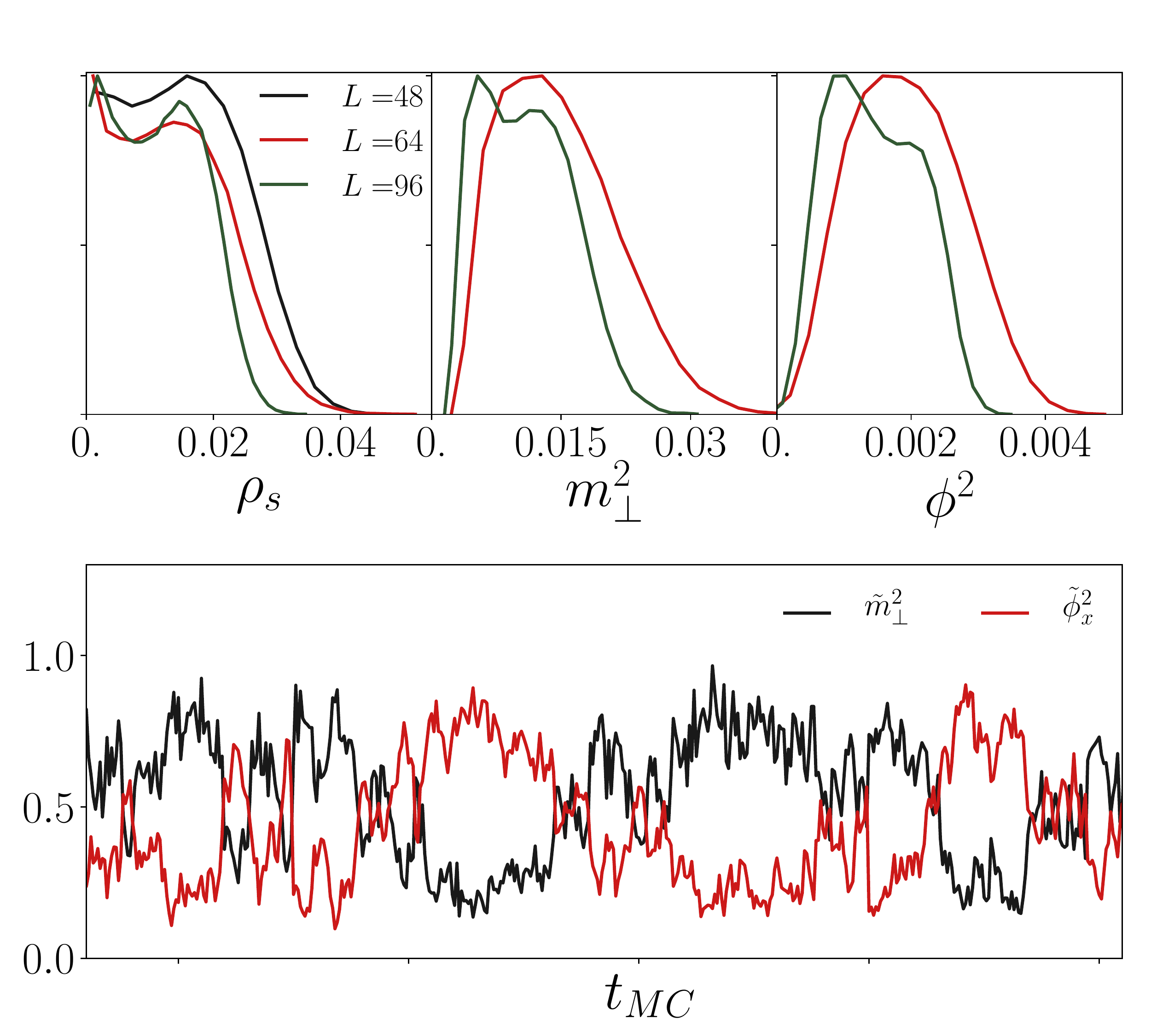}
 \caption{\label{fig:histlam50}Histograms (first row) and time series data for $L=96$ (second row) of observables close to the critical point ($g\approx12.1$) for $\lambda=0.5$. Here $\tilde{m}^2_\perp$ and $\tilde{\phi}_x^2$ are respectively  $m^2_\perp$ and $\phi_x^2$ normalized so that the maximum value is 1.0. This data has been collected for less than 5000 MC steps per bin. The histograms show double peaked behavior and time series data shows switching between two orders.}
 \end{figure}
To investigate the nature of the transition, we study the extrapolation of observables with system size at the critical point. For a continuous transition, all the observables described above ($\rho_s$, $\langle m^2 \rangle$, $\langle \phi^2 \rangle$), should go to zero at the critical point as $L\to\infty$. Figs \ref{fig:ordrprmgcextrpln} and \ref{fig:ordrprmgcextrpln2} shows values of $\langle m^2_{\perp} \rangle$ and $\langle \phi_x^2 \rangle$ at the crossing points of $\mathcal{R}_{m_\perp^2}$ and $\mathcal{R}_{\phi_x^2}$ at $L$ extrapolated to the infinite system size limit using two different fitting forms (as described in the caption). The extrapolated values of $\langle m_\perp^2 \rangle$ and $\langle \phi_x^2 \rangle$ are clearly finite for $\lambda=0.5$, $\lambda=0.75$ and $\lambda=0.85$.  Small positive extrapolated values of these quantities using both the fit forms can also be seen for $\lambda=0.95$. Fig.~\ref{fig:ordrprmcvslam} shows how the order parameter values at the transition point get progressively smaller on increasing $\lambda$, indicating a weakening of the first order nature. They tend to approach zero as $\lambda\rightarrow1$. Therefore we can argue that the first order transition continuously evolves on varying $\lambda$ to a second order transition at $\lambda=1.0$. 

The stiffness extracted from crossings of $L\rho_s$ for $L$ and $L/2$ in Fig \ref{fig:rhocfit}(a),(b) is plotted as a function of $1/L$ in \ref{fig:rhocfit}(c). $\rho_s$ clearly extrapolates to a finite value for $1/L \to 0$ for both $\lambda=0.5$ and $\lambda=0.75$. Fig.~\ref{fig:rhocfit2} shows the same analysis for $\lambda=0.85$ and $\lambda=0.95$. This points to a first order transition for $\lambda=0.5,0.75,0.85$ and $0.95$. For $\lambda=1$, on the other hand it is apparent from our data that is hard to argue for a finite order parameter for $\langle m_{\perp}^2 \rangle$ and $\langle \phi_x^2 \rangle$ from the data we have. A more thorough analysis of the $\lambda=1$ is available in Ref.~\onlinecite{Shao2016:science}. We note that these extrapolations become hard to do on very large system sizes because of ergodicity issues that we discuss below and that we argue arise fundamentally at first order transitions. 

\subsubsection{Histograms}
\label{subsubsec:histograms}
 
\begin{figure}[t]
    \centering
    \includegraphics[width=0.5\textwidth]{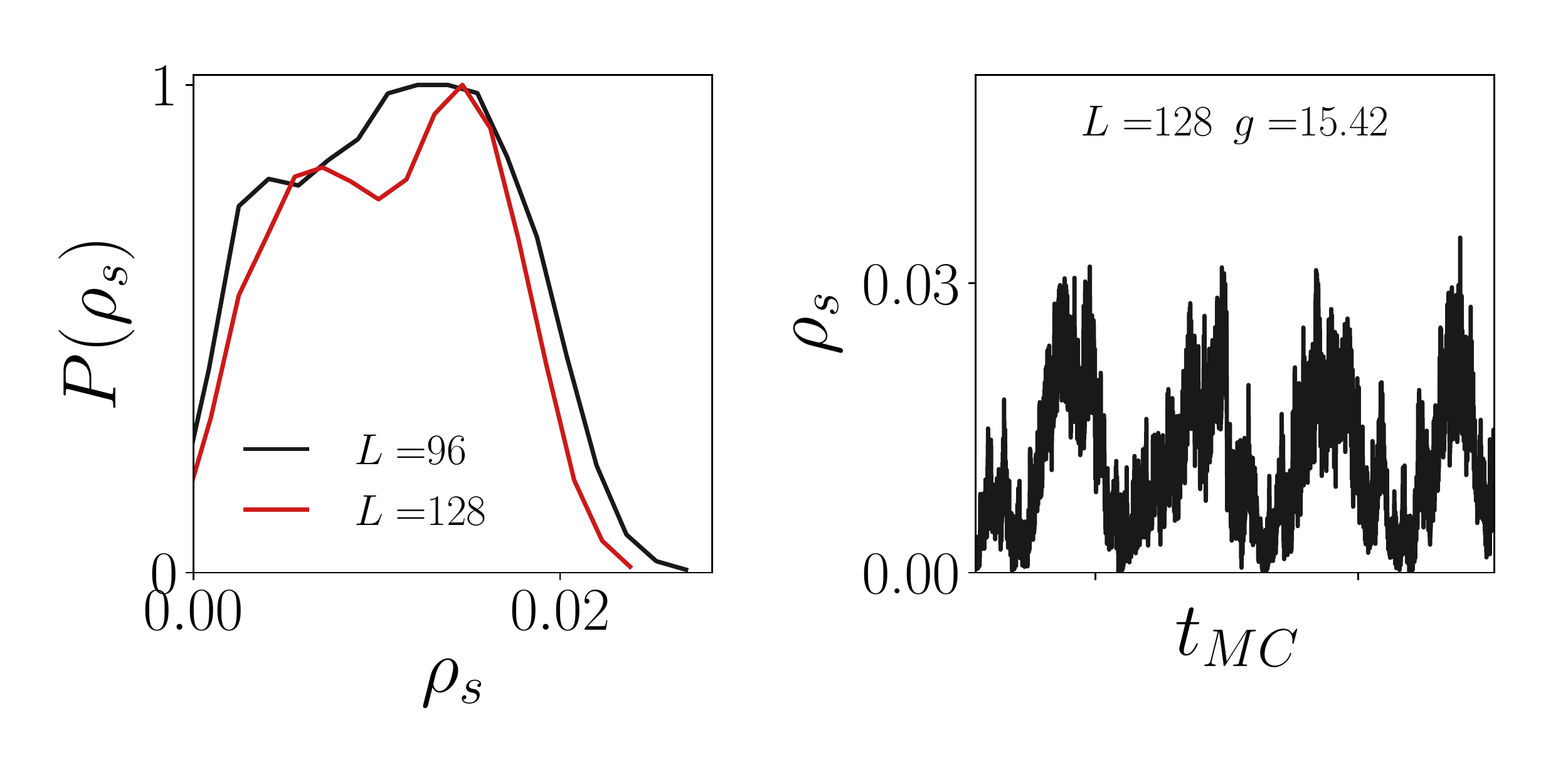}
    \caption{Histograms (left) and Monte Carlo histories (right) of $\rho_s$ for $\lambda=0.75$ near the critical point $g\approx15.4$. The histogram data has been collected for 1000 MC steps per bin. The double peak in the histograms that is barely visible for $L=96$ just starts to appear for $L=128$. Switching between the two values of $\rho_s$ depicted in the time series data also indicates first order behaviour.}
    \label{fig:histlam75}
\end{figure}

To further elucidate the nature of the transition we carefully study the histograms of observables near the critical point. Fig. \ref{fig:histlam50} shows the probability distributions of the QMC estimators for $\rho_s$, $m^2_{\perp}$ and $\phi_x^2$ at the transition for $\lambda=0.5$. There are clearly two peaks in the histograms of $\rho_s$ for $L=48, 64, 96$, one at 0 and the other at a finite value. This double peak feature is clearly noticeable in $m^2_{\perp}$ and $\phi_x^2$ only for $L=96$. The double peak gets more pronounced with system size which indicates that the first order behavior survives in the thermodynamic limit. The time series data shows switching between the two orders: one order parameter is finite when the other goes to 0, thus one order is present when the other is not. This is characteristic of a first order transition. This system exhibits clear first order behavior only for $L>64$, therefore we conclude that this transition is a weak first order transition. The first order nature of the transition is even weaker for $\lambda=0.75$, the double peak in the histograms of stiffness appears for $L>96$ as shown in Fig.  \ref{fig:histlam75}. We find no evidence of double peaked histograms for $\lambda=0.85$ and $0.95$ for the largest system sizes studied here. However, as explained before, we do find other evidence of first order behaviour in these two cases. There is also no evidence of double peaked histograms for $\lambda=1.0$ for the largest system size we have studied. This is consistent with the numerical findings in the past. \cite{sandvik2007:deconf,melko2008:fan,sandvik2010:deconf,harada2013:suNHeis}  Therefore we conclude that the transition is first order for $\lambda=0.5,0.75,0.85$ and $0.95$. The first order behaviour gets progressively weaker as $\lambda$ gets closer to 1, eventually disappearing at $\lambda=1$.
 
\section{Conclusions}
\label{subsec:conclusion}

We studied an interpolation of two previously known and well studied models, the J-Q model~\cite{sandvik2007:deconf} which hosts a continuous N\'eel-VBS transition and the easy-plane J-Q model~\cite{jon2016:easyplane} which hosts a first order superfluid-VBS transition. By studying the phase transition as a function of the parameter $\lambda$ that interpolates between the two limits, we found the phase diagram shown in Fig.~\ref{fig:phasediag}. Our main conclusion is that whenever the easy-plane anisotropy is present the transition is first order. All signs of discontinuity vanish  only at the symmetric point $\lambda=1$. This indicates that the easy-plane anisotropy is a relevant perturabation at the SU(2) symmetric deconfined critical point and results in a runaway flow to a first order transition.

We acknowledge helpful discussions with A. Sandvik. Partial financial support was received through NSF DMR-1611161 and Keith B Macadam Graduate Excellence Fellowship. Computing resources were obtained through NSF's XSEDE award TG-DMR-140061 and the DLX computer at the University of Kentucky.

\appendix

\section{Details of Numerical Simulations and Checks}
\label{sec:numsim}

\subsection{Lattice Hamiltonian}
\label{subsec:ham}
The Hamiltonian defined by Eq. \ref{eq:ham} is sign problem free on a bipartite lattice \footnote{The following unitary transformation: $S_x \rightarrow -S_x$ and $S_y \rightarrow -S_y$, on one of the sublattices of the bipartition, yields all negative off-diagonal elements for the Heisenberg exchange which is the sign problem free condition} and therefore we use the SSE QMC algorithm with directed loop updates to simulate it. The easy plane limit of this model ($\lambda=0$) has no diagonal terms. Hence, to make the model easier to simulate in this limit, we add a constant to the Hamiltonian to generate diagonal matrix elements.\cite{jon2016:easyplane} The easy plane part of the model defined in Eq. \ref{eq:epJQ} then becomes: 
\begin{equation}
    H^{JQ}_{ns}=J\sum_{\langle ij \rangle} (\tilde{P}_{ij}+\mathbbm{1}_{ij})  - Q \sum_{\langle ijkl \rangle} \tilde{P}_{ij}\tilde{P}_{kl}
    \label{eq:epJQ2}
\end{equation}
To make the loop update more convenient we treat all bonds as plaquettes by multiplying an identity to the adjacent bond, for e.g. the $P_{ij}$ operator in Eq. \ref{eq:singproj} gets replaced in the following way:
\begin{equation}
    P_{ij}=\frac{1}{N^{b}_{plaq}}\sum_{kl}P_{ij}.\mathbbm{1}_{kl}
    \label{eq:pijplaq}
\end{equation}
Here $\mathbbm{1}_{kl}$ is an identity operator, the sum in this equation is over all four site plaquettes $ijkl$ such that $kl$ is adjacent and parallel to $ij$. $N^{b}_{plaq}$ is number of plaquettes each bond is a part of, which is 2 in the square lattice case. After making these substitutions the full Hamiltonian described by Eq. \ref{eq:ham} becomes:

\begin{multline}
    H=\lambda\{\frac{J}{2}\sum_{ijkl}(P_{ij}.\mathbbm{1}_{kl}+\mathbbm{1}_{ij}.P_{kl})+Q\, P_{ij}.P_{kl})\}+ \\ (1-\lambda)\{\frac{J}{2}\sum_{ijkl}(\tilde{P}_{ij}.\mathbbm{1}_{kl}+\mathbbm{1}_{ij}.\tilde{P}_{kl}+\mathbbm{1}_{ij}.\mathbbm{1}_{kl})+Q\,\tilde{P}_{ij}.\tilde{P}_{kl}\}
    \label{eq:plqham}
\end{multline} \\

\begin{figure}[t]
    \centering
    \includegraphics[width=0.5\textwidth]{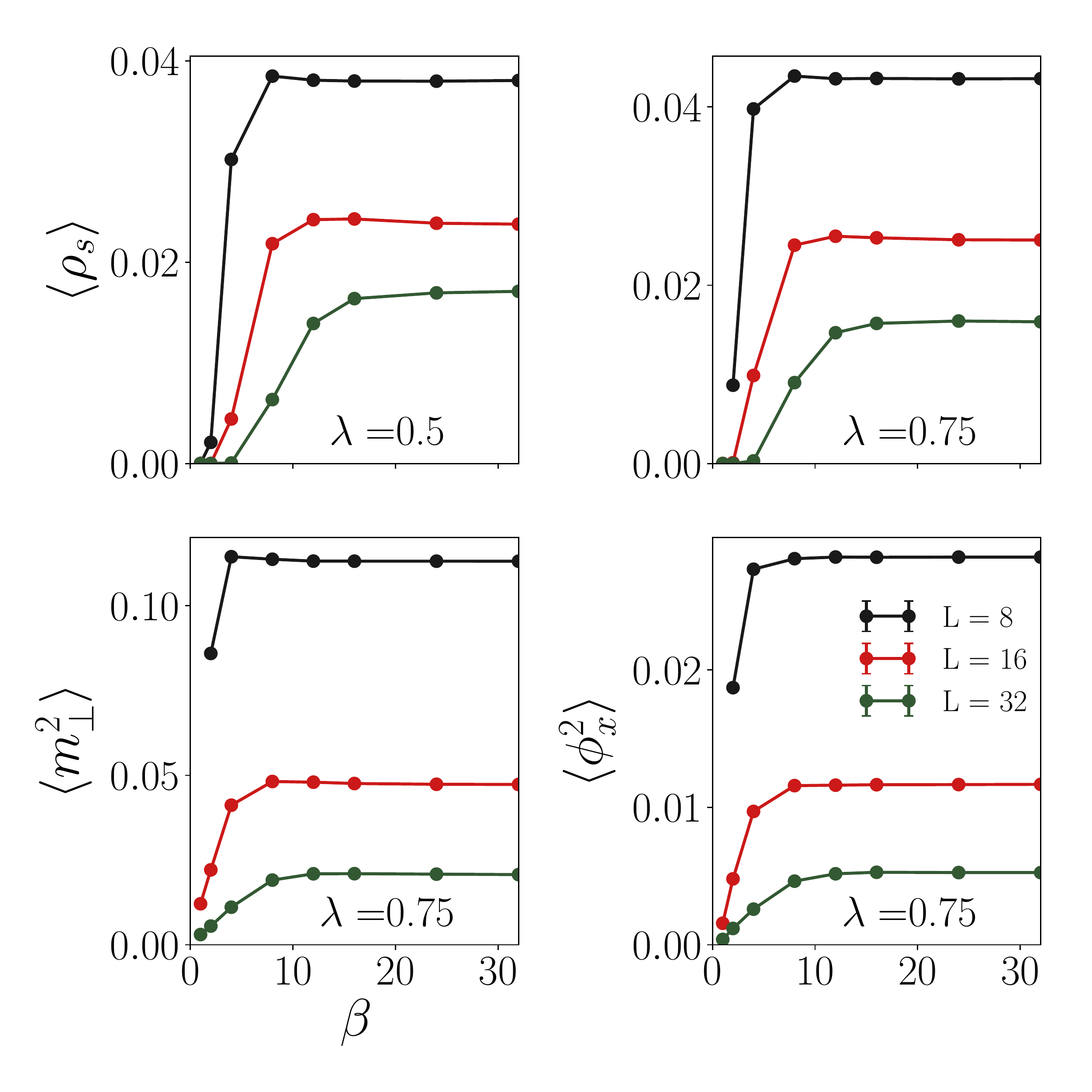}
    \caption{Observables $\rho_s$,$m^2_{\perp}$ and $\phi^2_{x}$ plotted vs $\beta$ can be seen to saturate on increasing $\beta$. This data has been taken at $g=12$ for $\lambda=0.5$ and $g=16$ for $\lambda=0.75$. These observables can be seen to saturate before $\beta=L$.}
    \label{fig:obsvsbeta}
\end{figure}

\begin{figure}
    \centering
    \includegraphics[width=0.4\textwidth]{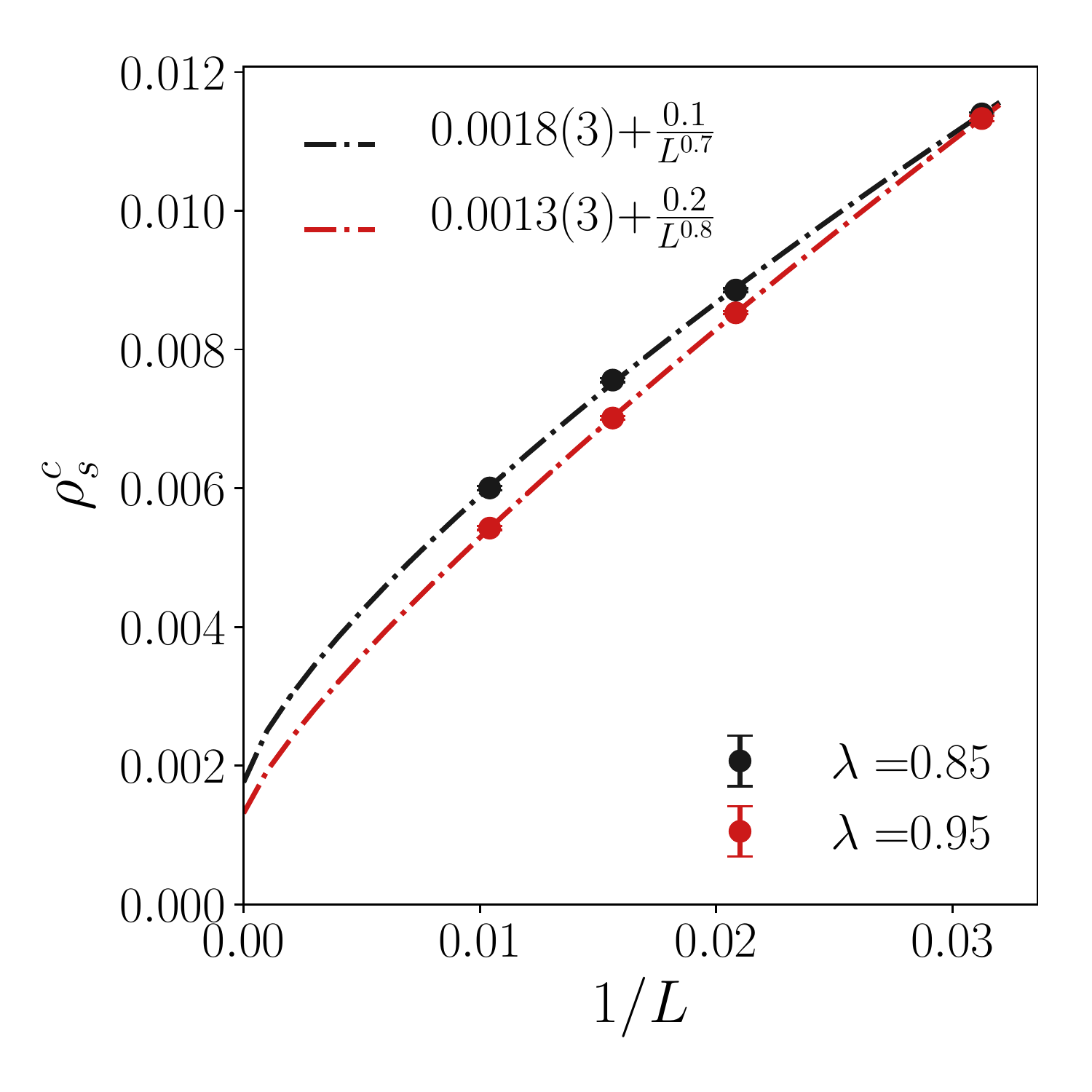}
    \caption{This analysis is the same one shown in fig~\ref{fig:rhocfit}(c) for $\lambda=0.85,0.95$. We have put these in a separate figure to avoid overcrowding the former. Here too, the stiffness extracted at the transition point goes to a finite value as $L \rightarrow \infty$.}
    \label{fig:rhocfit2}
\end{figure}{}

\begin{table}
  \caption{Comparison of ground state energy per unit site and spin stiffness values from QMC (@ $\beta = 4L$) with ED for $4\times4$ square lattice}
  \centering
  \begin{tabular}{c c c c c c c c c}
  \hline
   $\lambda$ & $\frac{Q}{J}$ & \textbf{$e^{qmc}$} & \textbf{$e^{exact}$} & \textbf{$\rho^{qmc}_s$} & \textbf{$\rho^{exact}_s$}\\
  \hline
  0.5 & 0.5 & -0.99366(6) & -0.99371 & 0.2737(1) & 0.2738\\
  0.5 & 1.0 & -0.96744(6) & -0.96746 & 0.2375(1) & 0.2374\\
  0.75 & 10.0 & -0.80616(6) & -0.80608  & 0.12102(7) & 0.12090\\
  0.75 & 18.0 & -0.76614(6) & -0.76613 & 0.11319(7)  & 0.11324\\
  \label{tab:obs}
  \end{tabular}
 \end{table}
 
 \begin{table}
  \caption{Comparison of order parameter from QMC (@ $\beta = 4 L$) with ED for $4\times4$ square lattice}
  \centering
  \begin{tabular}{c c c c c c c c c}
  \hline
  $\lambda$ & $\frac{Q}{J}$ & $ \langle m^{2}_{\perp} \rangle_{qmc}$ & $\langle m^2_{\perp} \rangle_{exact}$& $\langle \phi^2_x \rangle_{qmc}$ & $\langle \phi^2_x \rangle_{exact}$ \\
  \hline
 0.5 & 0.5 & 0.43721(6) & 0.43725 & 0.04416(2)  & 0.04414\\
 0.5 & 1.0 & 0.39560(5) & 0.39558  & 0.05884(3) & 0.05887\\
 0.5 & 15.0 & 0.26562(3) & 0.26560 & 0.04643(2) & 0.04642 \\
 0.75 & 10.0 & 0.27187(2)  & 0.27186 & 0.07704(3) & 0.07703\\
 0.75 & 18.0 & 0.26548(2)  & 0.26546 & 0.07317(2) & 0.7318\\
 \label{tab:corr}
 \end{tabular}
\end{table}
\subsection{Plaquette Operator}
\label{subsec:plaqop}
The plaquette operator, $\mathcal{Q}_x(\vec{r})$ in Eq. \ref{eq:plaqcorr} is the sum of all operators in the Hamiltonian acting on the plaquette at $\vec{r}$. Let $\vec{r}_{ijkl}$ be the position vector of the lower left site of the plaquette $ijkl$ 

\begin{multline}
    \mathcal{Q}_x(\vec{r}_{ijkl})=\frac{J}{2}\{\lambda\,(P_{ij}.\mathbbm{1}_{kl}+\mathbbm{1}_{ij}.P_{kl})\\+(1-\lambda)\,(\tilde{P}_{ij}.\mathbbm{1}_{kl}+\mathbbm{1}_{ij}.\tilde{P}_{kl}+\mathbbm{1}_{ij}.\mathbbm{1}_{kl})\}\\+Q\,\{\lambda \, P_{ij}.P_{kl}+(1-\lambda)\,\tilde{P}_{ij}.\tilde{P}_{kl}\}
    \label{eq:plaqop}
\end{multline} \\

\subsection{QMC vs ED}
Tables \ref{tab:obs} and \ref{tab:corr} show comparison of the groundstate energy per unit site ($e$), spin stiffness ($\rho_s$) and square of the order parameters, $\langle m_{\perp}^2 \rangle$ and $\langle \phi_x^2\rangle$, got from QMC and from exact diagonalization for $\lambda=0.5$ and $\lambda=0.75$ on $4\times4$ lattices. $\rho_s$, $m^2_{\perp}$ and $\phi_x^2$ are as defined in Sec. \ref{subsec:measurements}.

\subsection{Convergence to \texorpdfstring{$T=0$}{T=0}}
\label{subsec:obsvsbeta}
Fig. \ref{fig:obsvsbeta} shows the behaviour of the observables we have measured for $L\times L$ square lattices as a function of inverse temperature $\beta$. The measurements have been done close to the critical points ($g=12$ for $\lambda=0.5$ and $g=16$ for $\lambda=0.75$). These quantities can be seen to saturate to the $T=0$ value on increasing the value of $\beta$. The $\beta$ at which this saturation occurs depends on the system size $L$. As we increase the system size these values saturate to the value at $\beta=L$ faster, therefore we pick $\beta=L$ for our simulations.

\bibliographystyle{apsrev}
\bibliography{easy-plane.bib}

\end{document}